%% file: wof.tex
\documentclass{eptcs} % possibly tweak options
\usepackage{breakurl} % recommended by EPTCS style
\usepackage{proof}
\usepackage{xspace}
\usepackage{xcolor}
\usepackage{listings} % for listing code
\usepackage{wrapfig}

\input listing-macros

\newcommand{\blue}[1]{{\color[rgb]{0,0,1} #1}}

\newcommand{\tupp}[2]{\blue{\langle #1,}#2{\blue{\rangle}}}

\newcommand{\atac}{ACheck\xspace}
\newcommand{\lra}{\mathrel{\vdash}}
\newcommand{\seq}[2]{#1\lra #2}

\newcommand{\veep}{\vee^{\!+}}

\newcommand{\Nscr}{{\cal N}}
\newcommand{\Rscr}{{\cal R}}                                   % Used for an ambiguous rhs
\newcommand{\jUnf    }[4]{#1\mathbin\Uparrow#2\vdash#3\mathbin\Uparrow #4} % unfocused sequent
\newcommand{\jUnfamb }[3]{#1\mathbin\Uparrow#2\vdash#3 \Rscr}  % unfocused sequent
\newcommand{\jUnfGamb}[1]{\Gamma\mathbin\Uparrow#1\vdash \Rscr}% unf sequ with \Gamma
\newcommand{\jLf     }[3]{#1\Downarrow#2\vdash#3}              % left focused sequent
\newcommand{\jRf     }[2]{#1\vdash #2\Downarrow}               % right focused sequent
\newcommand{\bXi}[1]{\blue{\Xi_{#1} :\null}}

\newcommand{\storeClerk}[3]{\hbox{\sl store}_c(#1,#2,#3)}

\newcommand{\orExpert  }[3]{{\vee_e}(#1,#2,#3)}
\newcommand{\someExpert}[3]{\exists_e(#1,#2,#3)}

\newcommand{\decideExpert}[3]{\hbox{\sl decide}_e(#1,#2,#3)}

\title{Proof Outlines as Proof Certificates: A System Description}
\author{
  Roberto Blanco 
  \institute{Inria \& LIX, \'Ecole Polytechnique}
\and 
  Dale Miller
  \institute{Inria \& LIX, \'Ecole Polytechnique}
}
\providecommand{\event}{WoF'2015}
\def\titlerunning{Proof Outlines as Proof Certificates}
\def\authorrunning{R. Blanco \& D. Miller}

\begin{document}
\maketitle

\begin{abstract}\vskip -14pt
We apply the \emph{foundational proof certificate} (FPC) framework to
the problem of designing high-level outlines of proofs.
The FPC framework provides a means to formally define and check a wide
range of proof evidence.
A focused proof system is central to this framework and such a proof
system provides an interesting approach to proof reconstruction during
the process of proof checking (relying on an underlying logic
programming implementation).
Here, we illustrate how the FPC framework can be used to design proof
outlines and then to exploit proof checkers as a means for expanding
outlines into fully detailed proofs.
In order to validate this approach to proof outlines, we have built
the \atac system that allows us to take a sequence of theorems and
apply the proof outline ``do the obvious induction and close the proof
using previously proved lemmas''.
\end{abstract}

\section{Introduction}

Inference rules, as used in, say, Frege proofs (a.k.a. Hilbert proofs)
are usually greatly restricted by limitations of human psychology and
by what skeptics are willing to trust.
Typically, checking the application of inference rules involves
simple syntactic checks, such as deciding on whether or not two
premises have the structure $A$ and $A\supset B$ and the conclusion
has the structure $B$.
The introduction of automation into theorem proving has allowed us to
engineer inference steps that are more substantial and include both
computation and search.
In recent years, a number of proof theoretic results allow us to
extend that literature from being a study of minuscule inference rules
(such as \emph{modus ponens} or Gentzen's introduction rules) to a
study of large scale and formally defined ``synthetic'' inference
rules.
In this paper, we briefly describe the \atac system in which we build
and check \emph{proof outlines} as combinations of such synthetic rules.

Consider the following inductive specification of addition of natural numbers
\begin{lstlisting}[linerange=nat-end]
Kind      nat                 type.
Type      z                   nat.
Type      s                   nat -> nat.

Define nat : nat -> prop by
  nat z ;
  nat (s N) := nat N.

/* nat */
Define plus : nat -> nat -> nat -> prop by
  plus z N N ;
  plus (s N) M (s P) := plus N M P.
/* end */

/* plustotal */
Theorem plustotal :
  forall N, nat N -> forall M, nat M -> exists S, plus N M S.
/* end */
induction on 1. intros. case H1.
  search.
  apply IH to H3. apply H4 to H2. search.

/* plusdeterm */
Theorem plusdeterm : forall N, nat N -> forall M, nat M -> 
  forall S, plus N M S  -> forall T, plus N M T -> S = T.
/* end */
induction on 1. intros. case H1.
  case H3. case H4. search.
  case H3. case H4. apply IH to H5. apply H8 to H2. apply H9 to H6. apply H10 to H7. search.

/* plus0com */
Theorem plus0com : forall N, nat N ->  plus N z N.
/* end */
induction on 1. intros. case H1.
  search.
  apply IH to H2. search.

/* plusscom */
Theorem plusscom : forall M, nat M -> forall N, nat N -> 
                   forall P, plus M N P -> plus M (s N) (s P).
/* end */
induction on 1. intros. case H1.
  case H3. search.
  case H3. apply IH to H4. apply H6 to H2. apply H7 to H5. search.

/* pluscom */
Theorem pluscom : forall N, nat N -> forall M, nat M -> 
                  forall S, plus N M S  -> plus M N S.
/* end */
induction on 1. intros. case H1.
  case H3. apply plus0com to H2. search.
  case H3. apply IH to H4. apply H6 to H2. apply H7 to H5.
    apply plusscom to H2. apply H9 to H4. apply H10 to H8. search.
/* end */
\end{lstlisting} 
where {\tt z} and {\tt s} denote zero and successor, respectively.
(Examples will be displayed using the syntax of the Abella theorem prover
\cite{baelde14jfr}: this syntax should be familiar to users of other
systems, such as Coq.)
When this definition is introduced, we
should establish several properties immediately, e.g., that the
addition relation is determinate and total.
\begin{lstlisting}[linerange=plustotal-end]
Kind      nat                 type.
Type      z                   nat.
Type      s                   nat -> nat.

Define nat : nat -> prop by
  nat z ;
  nat (s N) := nat N.

/* nat */
Define plus : nat -> nat -> nat -> prop by
  plus z N N ;
  plus (s N) M (s P) := plus N M P.
/* end */

/* plustotal */
Theorem plustotal :
  forall N, nat N -> forall M, nat M -> exists S, plus N M S.
/* end */
induction on 1. intros. case H1.
  search.
  apply IH to H3. apply H4 to H2. search.

/* plusdeterm */
Theorem plusdeterm : forall N, nat N -> forall M, nat M -> 
  forall S, plus N M S  -> forall T, plus N M T -> S = T.
/* end */
induction on 1. intros. case H1.
  case H3. case H4. search.
  case H3. case H4. apply IH to H5. apply H8 to H2. apply H9 to H6. apply H10 to H7. search.

/* plus0com */
Theorem plus0com : forall N, nat N ->  plus N z N.
/* end */
induction on 1. intros. case H1.
  search.
  apply IH to H2. search.

/* plusscom */
Theorem plusscom : forall M, nat M -> forall N, nat N -> 
                   forall P, plus M N P -> plus M (s N) (s P).
/* end */
induction on 1. intros. case H1.
  case H3. search.
  case H3. apply IH to H4. apply H6 to H2. apply H7 to H5. search.

/* pluscom */
Theorem pluscom : forall N, nat N -> forall M, nat M -> 
                  forall S, plus N M S  -> plus M N S.
/* end */
induction on 1. intros. case H1.
  case H3. apply plus0com to H2. search.
  case H3. apply IH to H4. apply H6 to H2. apply H7 to H5.
    apply plusscom to H2. apply H9 to H4. apply H10 to H8. search.
/* end */
\end{lstlisting}
\begin{lstlisting}[linerange=plusdeterm-end]
Kind      nat                 type.
Type      z                   nat.
Type      s                   nat -> nat.

Define nat : nat -> prop by
  nat z ;
  nat (s N) := nat N.

/* nat */
Define plus : nat -> nat -> nat -> prop by
  plus z N N ;
  plus (s N) M (s P) := plus N M P.
/* end */

/* plustotal */
Theorem plustotal :
  forall N, nat N -> forall M, nat M -> exists S, plus N M S.
/* end */
induction on 1. intros. case H1.
  search.
  apply IH to H3. apply H4 to H2. search.

/* plusdeterm */
Theorem plusdeterm : forall N, nat N -> forall M, nat M -> 
  forall S, plus N M S  -> forall T, plus N M T -> S = T.
/* end */
induction on 1. intros. case H1.
  case H3. case H4. search.
  case H3. case H4. apply IH to H5. apply H8 to H2. apply H9 to H6. apply H10 to H7. search.

/* plus0com */
Theorem plus0com : forall N, nat N ->  plus N z N.
/* end */
induction on 1. intros. case H1.
  search.
  apply IH to H2. search.

/* plusscom */
Theorem plusscom : forall M, nat M -> forall N, nat N -> 
                   forall P, plus M N P -> plus M (s N) (s P).
/* end */
induction on 1. intros. case H1.
  case H3. search.
  case H3. apply IH to H4. apply H6 to H2. apply H7 to H5. search.

/* pluscom */
Theorem pluscom : forall N, nat N -> forall M, nat M -> 
                  forall S, plus N M S  -> plus M N S.
/* end */
induction on 1. intros. case H1.
  case H3. apply plus0com to H2. search.
  case H3. apply IH to H4. apply H6 to H2. apply H7 to H5.
    apply plusscom to H2. apply H9 to H4. apply H10 to H8. search.
/* end */
\end{lstlisting} 
Anyone familiar with proving such theorems knows that their proofs are
simple: basically, the obvious induction leads quickly to a final proof.
Of course, if we wish to prove more results about addition, one may
need to invent and prove some lemma before simple inductions will
work.
For example, proving the commutativity of addition makes use of 
two additional lemmas.
\begin{lstlisting}[linerange=plus0com-end]
Kind      nat                 type.
Type      z                   nat.
Type      s                   nat -> nat.

Define nat : nat -> prop by
  nat z ;
  nat (s N) := nat N.

/* nat */
Define plus : nat -> nat -> nat -> prop by
  plus z N N ;
  plus (s N) M (s P) := plus N M P.
/* end */

/* plustotal */
Theorem plustotal :
  forall N, nat N -> forall M, nat M -> exists S, plus N M S.
/* end */
induction on 1. intros. case H1.
  search.
  apply IH to H3. apply H4 to H2. search.

/* plusdeterm */
Theorem plusdeterm : forall N, nat N -> forall M, nat M -> 
  forall S, plus N M S  -> forall T, plus N M T -> S = T.
/* end */
induction on 1. intros. case H1.
  case H3. case H4. search.
  case H3. case H4. apply IH to H5. apply H8 to H2. apply H9 to H6. apply H10 to H7. search.

/* plus0com */
Theorem plus0com : forall N, nat N ->  plus N z N.
/* end */
induction on 1. intros. case H1.
  search.
  apply IH to H2. search.

/* plusscom */
Theorem plusscom : forall M, nat M -> forall N, nat N -> 
                   forall P, plus M N P -> plus M (s N) (s P).
/* end */
induction on 1. intros. case H1.
  case H3. search.
  case H3. apply IH to H4. apply H6 to H2. apply H7 to H5. search.

/* pluscom */
Theorem pluscom : forall N, nat N -> forall M, nat M -> 
                  forall S, plus N M S  -> plus M N S.
/* end */
induction on 1. intros. case H1.
  case H3. apply plus0com to H2. search.
  case H3. apply IH to H4. apply H6 to H2. apply H7 to H5.
    apply plusscom to H2. apply H9 to H4. apply H10 to H8. search.
/* end */
\end{lstlisting}
\begin{lstlisting}[linerange=plusscom-end]
Kind      nat                 type.
Type      z                   nat.
Type      s                   nat -> nat.

Define nat : nat -> prop by
  nat z ;
  nat (s N) := nat N.

/* nat */
Define plus : nat -> nat -> nat -> prop by
  plus z N N ;
  plus (s N) M (s P) := plus N M P.
/* end */

/* plustotal */
Theorem plustotal :
  forall N, nat N -> forall M, nat M -> exists S, plus N M S.
/* end */
induction on 1. intros. case H1.
  search.
  apply IH to H3. apply H4 to H2. search.

/* plusdeterm */
Theorem plusdeterm : forall N, nat N -> forall M, nat M -> 
  forall S, plus N M S  -> forall T, plus N M T -> S = T.
/* end */
induction on 1. intros. case H1.
  case H3. case H4. search.
  case H3. case H4. apply IH to H5. apply H8 to H2. apply H9 to H6. apply H10 to H7. search.

/* plus0com */
Theorem plus0com : forall N, nat N ->  plus N z N.
/* end */
induction on 1. intros. case H1.
  search.
  apply IH to H2. search.

/* plusscom */
Theorem plusscom : forall M, nat M -> forall N, nat N -> 
                   forall P, plus M N P -> plus M (s N) (s P).
/* end */
induction on 1. intros. case H1.
  case H3. search.
  case H3. apply IH to H4. apply H6 to H2. apply H7 to H5. search.

/* pluscom */
Theorem pluscom : forall N, nat N -> forall M, nat M -> 
                  forall S, plus N M S  -> plus M N S.
/* end */
induction on 1. intros. case H1.
  case H3. apply plus0com to H2. search.
  case H3. apply IH to H4. apply H6 to H2. apply H7 to H5.
    apply plusscom to H2. apply H9 to H4. apply H10 to H8. search.
/* end */
\end{lstlisting}
\begin{lstlisting}[linerange=pluscom-end]
Kind      nat                 type.
Type      z                   nat.
Type      s                   nat -> nat.

Define nat : nat -> prop by
  nat z ;
  nat (s N) := nat N.

/* nat */
Define plus : nat -> nat -> nat -> prop by
  plus z N N ;
  plus (s N) M (s P) := plus N M P.
/* end */

/* plustotal */
Theorem plustotal :
  forall N, nat N -> forall M, nat M -> exists S, plus N M S.
/* end */
induction on 1. intros. case H1.
  search.
  apply IH to H3. apply H4 to H2. search.

/* plusdeterm */
Theorem plusdeterm : forall N, nat N -> forall M, nat M -> 
  forall S, plus N M S  -> forall T, plus N M T -> S = T.
/* end */
induction on 1. intros. case H1.
  case H3. case H4. search.
  case H3. case H4. apply IH to H5. apply H8 to H2. apply H9 to H6. apply H10 to H7. search.

/* plus0com */
Theorem plus0com : forall N, nat N ->  plus N z N.
/* end */
induction on 1. intros. case H1.
  search.
  apply IH to H2. search.

/* plusscom */
Theorem plusscom : forall M, nat M -> forall N, nat N -> 
                   forall P, plus M N P -> plus M (s N) (s P).
/* end */
induction on 1. intros. case H1.
  case H3. search.
  case H3. apply IH to H4. apply H6 to H2. apply H7 to H5. search.

/* pluscom */
Theorem pluscom : forall N, nat N -> forall M, nat M -> 
                  forall S, plus N M S  -> plus M N S.
/* end */
induction on 1. intros. case H1.
  case H3. apply plus0com to H2. search.
  case H3. apply IH to H4. apply H6 to H2. apply H7 to H5.
    apply plusscom to H2. apply H9 to H4. apply H10 to H8. search.
/* end */
\end{lstlisting}
These three lemmas have the same high-level proof outline:
use the obvious induction invariant, apply some previously proved
lemmas and the inductive hypothesis, and deal with any remaining
case analysis.

The fact that many theorems can be proved by using
induction-lemmas-cases is well-known and built into existing theorem
provers.
For example, the waterfall model of the Boyer-Moore prover
\cite{boyer79} proves such theorems in a similar fashion (but for
inductive definitions of functions).
Twelf \cite{pfenning99cade} can often prove that some
relations are total and functional using a series of similar steps
\cite{schurmann03tphols}.
The tactics and tacticals of LCF have been used to implement procedures
that attempt to find proofs using these steps \cite{wilson10coq}.
Finally, TAC \cite{baelde10ijcar} attempts to follow
such a procedure as well but in a rather fixed and inflexible fashion.

In this paper, we present an approach to describing the simple rules
that can prove a given lemma based on previously proved lemmas.
Specifically, we define proof certificates that describe the
structure of a proof outline that we expect and then we
run a proof checker on that certificate to see if the certificate
can be elaborated into a full proof of the lemma.

\section{A focused proof system}
\label{sec:intro}

\begin{wrapfigure}{r}{50mm}
\vspace{-.8cm}
\[
\infer{\seq{\Gamma}{B_1\lor B_2}}{\seq{\Gamma}{B_i}} 
\qquad
\infer{\seq{\Gamma}{\exists x.B}}{\seq{\Gamma}{B[t/x]}} 
\]
\caption{From the LJ calculus}
\label{fig:two unfocused}
\vspace{-.5cm}
\end{wrapfigure}

Consider the two introduction rules in Figure~\ref{fig:two unfocused}.
If one attempts to prove sequents by reading these rules from
conclusion to premises, then these rules need either information from
some external source (e.g., an oracle providing the $i\in\{1,2\}$ or
the term $t$) or some implementation support for non-determinism (e.g.,
unification and backtracking search). 

\begin{wrapfigure}{l}{50mm}
\vspace{-.5cm}
\[
\infer{\jRf{\Gamma}{B_1\lor B_2}}{\jRf{\Gamma}{B_i}} 
\qquad
\infer{\jRf{\Gamma}{\exists x.B}}{\jRf{\Gamma}{B[t/x]}} 
\]
\vspace{-.8cm}
\caption{Focusing annotations}
\vspace{-.5cm}
\label{fig:two up}
\end{wrapfigure}

It is meaningless to use Gentzen's sequent calculus to
directly support proof automation.  Consider, for example,
attempting to prove the sequent
$
\seq{\Gamma}{\exists x\exists y [(p~x~y)\lor ((q~x~y)\lor (r~x~y))]},
$
where $\Gamma$ contains, say, a hundred formulas.  The search for a
(cut-free) proof of this sequent can confront the need to choose from among a
hundred-and-one introduction rules.  If we choose the right-side
introduction rule, we will then be left with, again, a hundred-and-one
introduction rules to apply to the premise.  Thus, reducing this
sequent to, say, $\seq{\Gamma}{(q~t~s)}$ requires picking one path of
choices in a space of $101^4$ choices.

Focused proof systems address this explosion by organizing
rules into two phases.  For example, the
rules in Figure~\ref{fig:two unfocused} are written instead as
Figure~\ref{fig:two up}. 
Here, the formula on which one is introducing a connective is marked
with the $\Downarrow$: as a result,
reducing the sequent
$
\jRf{\Gamma}{\exists x\exists y [(p~x~y)\lor ((q~x~y)\lor (r~x~y))]}
$
to $\jRf{\Gamma}{(q~t~s)}$ involves only those choices related to the
formula marked for focus: no interleaving of other choices needs to be
considered.

While the $\Downarrow$ phase involves rules that may not be
invertible, the $\Uparrow$ phase involves rules that must be
invertible.  For example, the left rules for $\lor$
and $\exists$ are invertible and their introduction rule is listed as
\[
  \infer{\jUnfGamb{B_1\veep B_2,\Theta}} %[\veep_l]
                 {\jUnfGamb{B_1,\Theta}\quad \jUnfGamb{B_2,\Theta}}
\qquad
  \infer{\jUnfGamb{\exists x.B, \Theta}}{\jUnfGamb{[y/x]B,\Theta}} %[\exists_l]
\]
These rules need no external information (in particular, any new
variable $y$ will work in the $\exists$ introduction rule).  In these
last two rules, the zone between $\Uparrow$ and $\vdash$ contains a
\emph{list} of formulas.  When there are no more invertible rules that
can be applied to that first formula, that formula is moved to (i.e.,
stored in) the zone written as $\Gamma$, using the following
\emph{store-left} rule 
\[
  \infer[S_l.]{\jUnfamb{\Gamma}{C,\Theta}{\null}}{\jUnfamb{C,\Gamma}{\Theta}{\null}}
\]
Finally, when the zone between the $\Uparrow$ and the $\vdash$ is
empty (i.e., all invertible inference rules have been completed),
it is time to select a (possibly non-invertible) introduction rule to
attempt.  For that, we have the two \emph{decide} rules
\[
  \vcenter{\infer[D_l]{\jUnf{\Gamma,N}{\cdot}{\cdot}{E}}{\jLf{\Gamma,N}{N}{E}}}
  \qquad\hbox{and}\qquad
  \vcenter{\infer[D_r.]{\jUnf{\Gamma}{\cdot}{\cdot}{P}}{\jRf{\Gamma}{P}}}
\]

Although we cannot show all focused inference rules, we will
present those that deal with the least fixed point operator.  Formally
speaking, when we 
define a predicate, such as {\tt plus} in the previous section, we are
actually naming a least fixed point expression.  In the case of {\tt
  plus}, that expression is
\[
\mu\lambda P\lambda n\lambda m\lambda p.
  (n = z \wedge m = p) \vee
  \exists n' \exists p'.[n = (s~n') \wedge p = (s~p') \wedge (P~n'~m~p')].
\]
For the treatment of least fixed points, we follow the
$\mu LJ$ proof system and its focused variant 
\cite{baelde12tocl,baelde07lpar}.  The treatment of least fixed
point expressions in the $\Uparrow$ phase and the $\Downarrow$ phase
is given by the three rules
\[
\infer{\jUnfamb{\Gamma}{\mu B\bar t,\Theta}{\null}}
      {\jUnfamb{\mu B\bar t,\Gamma}{\Theta}{\null}}
\qquad
\infer{\Gamma\mathbin\Uparrow\mu B\bar t,\Theta\vdash N}
      {\Gamma\mathbin\Uparrow S\bar t,\Theta\vdash N\quad
             \mathbin\Uparrow B S \bar x\vdash S\bar x}
\qquad
\infer{\jRf{\Gamma}{\mu B\bar t}}{\jRf{\Gamma}{B(\mu B)\bar t}}
\]
Notice that the right introduction rule is just an unfolding of the
fixed point.  There are two ways to treat the least fixed point on the
left: one can either perform a store operation or one can do an
induction using, in this case, the invariant $S$.  The right premise of
the induction rule shows that $S$ is a prefixed point (i.e., $B
S\subseteq S$).  
In general, supplying an invariant can be tedious so we shall also
identify two admissible rules for unfolding (also on the left) and
\emph{obvious induction}, meaning that the invariant to use is nothing
more than the immediately surrounding sequent as the invariant $S$.
In the case of the obvious induction, the left premise sequent will be
trivial.

\section{Foundational proof certificates}
\label{sec:fpc}

The main idea behind using the \emph{foundational proof certificate}
(FPC) \cite{chihani13cade} approach to defining proof evidence is that
theorem provers output their proof evidence to some persistent memory
(e.g., a file) and that independent and trustable proof checkers
examine this evidence for validity.
In order for such a scheme to work, the semantics of the output proof
evidence must be formally defined.
The FPC framework provides ways to formally define such proof
semantics which is also executable when interpreted on top of a
suitable logic engine.

There are four key ingredients to providing such a formal definition
and they are all described via their relationship to focused proof
systems.
In fact, consider the following \emph{augmented} versions of inference
rules we have seen in the previous section.
\[
\infer{\bXi0 \jRf{\Gamma}{B_1\lor B_2}}
      {\bXi1 \jRf{\Gamma}{B_i}\quad \blue{\orExpert{\Xi_0}{\Xi_1}{i}}}
\qquad
\infer{\bXi0 \jRf{\Gamma}{\exists x.B}}
      {\bXi1 \jRf{\Gamma}{B[t/x]}\quad\blue{\someExpert{\Xi_0}{\Xi_1}{t}}}
\]
These two augmented rules contain two of the four ingredients of an
FPC: the schema variable $\Xi$ ranges over terms that denote the
actual proof evidence comprising a certificate.
The additional premises involve \emph{experts} which are predicates that
relate the concluding certificate $\Xi_0$ to a continuation
certificate $\Xi_1$ and some additional information.
The expert predicate for the disjunction can provide an indication of
which disjunct to pick and the expert for the existential quantifier
can provide an indication of which instance of the quantifier to use.
Presumably, these expert predicates are capable of digging into a
certificate and extracting such information.
It is not required, however, for an expert to definitively extract
such information.
For example, the disjunction expert might guess both 1 and 2 for $i$:
the proof checker will thus need to handle such non-determinism during
the checking of certificates.
\begin{wrapfigure}{r}{70mm}
\vspace{-.2cm}
\[
  \infer[S_l]{\bXi0 \jUnfamb{\Gamma}{C,\Theta}{\null}}
             {\bXi1 \jUnfamb{\tupp{l}{C},\Gamma}{\Theta}{\null}\quad
              \blue{\storeClerk{\Xi_0}{\Xi_1}{l}}}
\]
\[
  \infer[D_l]{\bXi0 \jUnf{\Gamma,\tupp{l}{N}}{\cdot}{\cdot}{E}}
             {\bXi1 \jLf{\Gamma,\tupp{l}{N}}{N}{E}\quad
              \blue{\decideExpert{\Xi_0}{\Xi_1}{l}}}
\]
\caption{Two augmented rules}
\vspace{-.5cm}
\label{fig:augmented}
\end{wrapfigure}

Another ingredient of an FPC is illustrated by the augmented 
inference rules in Figure~\ref{fig:augmented}.  
The store-left ($S_l$) inference rule is augmented
with an extra premise that invokes a \emph{clerk} predicate which is
responsible for computing an index $l$ that is associated to the
stored formula (here, $C$).  The augmented decide-left ($D_l$) rule is
given an extra premise that uses an expert predicate: that premise 
can be used to compute the index of the formula that is to be selected for focus.
The indexing mechanism does not need to be functional (i.e., different
formulas can have the same index) in which case the decide rule must
also be interpreted as non-deterministic.  In earlier work
\cite{chihani13cade}, indexes have been identified with
structures as diverse as formula occurrences and de Bruijn numerals.
In this paper, the only role planned by indexes will be as the names
given to lemmas and to hypotheses (i.e., formulas that are stored on
the left using the $S_l$ inference rule).

As indicated above, there are essentially three operations that we can
perform to treat a least fixed point formula in the left-hand
context.  In fact, we shall expand these into the following four
augmented inference rules.
\begin{enumerate}

\item The fixed point can be ``frozen'' in the sense that the
  store-left ($S_l$) rule is applied to it.  As a result of such a
  store operation, the stored occurrence of the fixed point will never
  be unfolded and will not be the site of an induction.  Such a frozen
  fixed point can only be used later in proof construction within an
  instance of the initial rule.

\item The fixed point can be unfolded just as it can be on the
  right-hand side of the sequent.  (Unfolding on the left is a
  consequence of the more general induction rule.)  The following
  augmented inference rule can be used to control when such a fixed
  point is unfolded.
\[
  \infer
    {\blue{\Xi_0        :}~\Nscr\Uparrow \mu{}B\,\bar t,  \Gamma\vdash\Rscr}
    {\blue{\Xi_1        :}~\Nscr\Uparrow B(\mu B)\,\bar t,\Gamma\vdash\Rscr
     \qquad
     \blue{\hbox{\sl unfold}(\Xi_0,\Xi_1)}}
\]

\item The most substantial inference rule related to the least fixed
  point is the induction rule.  In its most general form, this
  inference rule involves proving premises that involve an invariant.
  A proof certificate term $\Xi_0$ could include such an invariant
  explicitly and the following augmented rule could be used to extract
  and use that invariant.
\[
  \infer
    {\blue{\Xi_0        :}~\Nscr\Uparrow \mu{}B\,\bar t,\Gamma \vdash \Rscr}
    {\blue{\Xi_1        :}~\Nscr\Uparrow S\,\bar t,\Gamma \vdash \Rscr
     \qquad
     \blue{\Xi_2\,\bar y:}~\Nscr\Uparrow B\,S\,\bar y \vdash S\,\bar y
     \qquad
     \blue{\hbox{\sl ind}(\Xi_0,\Xi_1,\Xi_2,S)}}
\]

\item In general, it appears that invariants are often complex, large,
  and tedious structures to build and use.  Thus, it is most likely
  that we need to develop a number of techniques by which invariants
  are not built directly but are rather implied by alternative
  reasoning principles.  For example, the Abella theorem prover
  \cite{baelde14jfr}, allows the user to do induction not by
  explicitly entering an invariant but rather by performing a certain
  kind of guarded, circular reasoning.  In the context of this paper,
  we consider, however, only one approach to specifying induction and
  that involves taking the sequent $\Nscr\Uparrow \mu{}B\,\bar
  t,\Gamma \vdash \Rscr$ and abstracting out the fixed point expression
  to yield the ``obvious'' invariant $\hat S$ so that the premise 
  $\Nscr\Uparrow S\,\bar t,\Gamma \vdash \Rscr$ has an easy proof.
  As a result, only the second premise related to the induction rule
  needs to be proved.  The following augmented rule is used to
  generate and check whether or not the obvious induction invariant
  can be used.
\[
  \infer
    {\blue{\Xi_0       :}~\Nscr\Uparrow \mu{}B\,\bar t,\Gamma \vdash \Rscr}
    {\blue{\Xi_1\,\bar y:}~\Nscr\Uparrow B\,\hat S\,\bar y \vdash\hat S\,\bar y
     \qquad
     \blue{\hbox{\sl obvious}(\Xi_0,\Xi_1)}}
\]
\end{enumerate}

An FPC definition is then a collection of the following: 
$(i)$ the term constructors for the term encoding proof evidence (the
$\Xi$ schema variable above), 
$(ii)$ the constructors for building indexes, and
$(iii)$ the definitions of predicates for describing how the clerks and
experts behave in different inference rules.
Given these definitions, we then check whether or not a sequent of the
form $\blue{\Xi : \null}\seq{\Gamma}{B}$ is provable.  This latter
check can be done using a logic programming engine since such an
engine should support both unification and backtracking search
(thereby allowing one to have a trade-off between large certificates
with a great deal of explicit information and small certificates
where details are elided and reconstructed as needed).
In our own work, we have used both $\lambda$Prolog
\cite{miller12proghol} and Bedwyr~\cite{Bedwyr} as that logic-based
engine.

\section{Certificate design}
\label{sec:design}

Imagine telling a colleague ``The proof of this theorem follows by a
simple induction and the three lemmas we just proved.''
You may or may not be correct in such an assertion since (a) the
proposed theorem may not be provable and (b) the simple proof you
describe may not exist.
In any case, it is clear that there is a rather simple, high-level
algorithm to follow that will search for such a proof.
In this section, we translate that algorithm into an FPC.  

Following the paradigm of focused proof systems for first-order logic,
there is a clear, high-level outline to follow for doing proof search
for cut-free proofs: first do all invertible inference rules and then,
select a formula on which to do a series of non-invertible choices.
This latter phase ends when one encounters invertible inference rules
again or the proof ends.
In the setting we describe here, there are two significant
complicating features with which to be concerned.

\newcommand{\parag}[1]{\smallskip\noindent\emph{#1}\ }

\parag{Treating the induction rule.}
The invertible ($\Uparrow$) phase is generally a place where no
important choices in the search for a proof appear.
When dealing with a formula that is a fixed point, however, this is no
longer true.
As described in the previous section, we treat that fixed point
expression either by freezing (see also \cite{baelde12tocl}),
unfolding, or using an explicitly supplied invariant or the
``obvious'' induction.

\parag{Lemmas must be invoked.}
% The application of lemmas into a proof outline is critical to the
% kind of linear proof development we have in mind.  
Although the focusing framework can work with any provable formula as
a lemma, we shall only consider lemmas that are Horn clauses (for
example, the lemmas at the end of Section~\ref{sec:intro}).
Consider applying a lemma of the form $\forall\bar x[A_1\supset
  A_2\supset A_3]$ in proving the sequent $\seq{\Gamma}{E}$.
Since the formulas
$A_1$, $A_2$, and $A_3$ are polarized positively, we can
design the proof outline (simply by only authorizing fixed points to
be frozen during this part of the proof) so that
$\jLf{\Gamma}{\forall\bar x[A_1\supset A_2\supset A_3]}{E}$ is
provable if and only if there is a substitution $\theta$ for the
variables in the list of variables $\bar x$ such that $\theta A_1$ and
$\theta A_2$ are in $\Gamma$ and the sequent $\seq{\Gamma, \theta
  A_3}{E}$ is provable.  The application of such a lemma is then
seen as forward chaining: if the context $\Gamma$
contains two atoms (frozen fixed points) then add a third. 
% Augmenting contexts in this way is critical for eventually enabling
% obvious inductions to succeed in completing a proof.  In this way, the
% focused proof system can easily be used to apply lemmas.

The main issue that a certificate-as-proof-outline therefore needs to
provide is some indication of what lemmas should be used during the
construction of a proof.  
The following collections of supporting
lemmas---starting from the least explicit to the most explicit---have been
tested within our framework: 
$(i)$ a bound on the number of lemmas that can be used to finish the proof;
$(ii)$ a list of possible lemmas to use in finishing the proof; and $(iii)$
a tree of lemmas, indicating which lemmas are applied following the
conjunctive structure of the remaining proof.

\section{The proof checker}
\label{sec:checker}

A direct and natural way to implement the FPC approach to proof
checking is to use a logic programming language:
by turning the augmented inference rules sideways, one gets logic
programming clauses.
In this way, the rule's conclusion becomes the head of the clause and
the rule's premises become the body of the clause.
When proof checking in (either classical or intuitionistic)
first-order logic without fixed points, the $\lambda$Prolog
programming language \cite{miller12proghol} is a good choice since its
treatment of bindings allows for elegant and effective implementations
of first-order quantification in formulas and of eigenvariables in
proofs \cite{chihani13cade}.
When the logic itself contains fixed points, as is the case in this
paper, $\lambda$Prolog is no longer a good choice: instead, a stronger
logic that incorporates aspects of the \emph{closed world assumption}
is needed.
%
% Our \atac system employs the Bedwyr model checking system
%
%In particular, the \atac system has three parts.  
%The first is a theorem prover: we have used Abella since
%it was easy to modify for exporting theorems and certificates.
%The second is the Bedwyr model checking system \cite{Bedwyr} and 
%the third, new component, is the Bedwyr code underlying this
%particular paper.  That code, called FPCcheck, is located at
%\verb+https://github.com/proofcert/fpccheck+.
%The documentation at that address explains where to find Bedwyr and
%our modified Abella system.
In particular, the \atac system has two parts.  
The first is a theorem prover; we have used Abella since
it was easy to modify it for exporting theorems and certificates for use by
the second part.  The second part is the proof checker, FPCcheck, that verifies
the output of a session of the theorem prover, suitably translated.
This checker is written in the Bedwyr model checking system \cite{Bedwyr} and 
is the new component underlying this
particular paper: its code is available from
\verb+https://github.com/proofcert/fpccheck+.
The documentation at that address explains where to find Bedwyr and
our modified Abella system.

To illustrate here the kinds of examples available on the web page,
the Abella theory files can have a \verb+ship+ command that is
followed by a string describing a certificate to use to prove the
proposed theorem: the checking of this certificate is shipped to the
Bedwyr-based kernel for checking.  In this particular case, the \verb+induction+
certificate constructor is given three arguments: the first is the
maximal number of decides that can be used in the proof and the second
and third are bounds on the number of unfoldings in the $\Uparrow$ and
$\Downarrow$ phases respectively.
\begin{lstlisting}
Theorem plus0com : forall N, is_nat N ->  plus N zero N.
   ship "(induction 1 0 1)".
Theorem plusscom : forall M, is_nat M -> forall N, is_nat N -> 
             forall P, plus M N P -> plus M (succ N) (succ P).
   ship "(induction 1 0 1)".
Theorem pluscom : forall N, is_nat N -> forall M, is_nat M -> 
                  forall S, plus N M S  -> plus M N S.
   ship "(induction 2 1 0)".
\end{lstlisting}
The bound on the number of decide rules (first argument) is also a bound
on the number of lemmas that can be used on any given branch of the proof.

\section{System architecture}

Our system for producing and checking proof outlines follows a simple
linear work-flow: first, state a theorem and obtain a proof outline,
next, attempt to check the theorem with the outline given as a proof
certificate.  We have structured this process in three computation
steps, involving a theorem prover, a translator, and a proof
checker. Their roles are given here.

\parag{The theorem prover.} An existing theorem prover is principally
the source of the concrete syntax of definitions and theorems.  It
may not be directly involved in the work-flow, particularly if the
proof language is extended to support \verb+ship+ tactics that enable
the omission of detailed proof scripts. At the end of the phase a \emph{theorem
  file} with all relevant definitions, theorem statements and proof
scripts is obtained.

\parag{The translator.} For each theorem prover, we need to furnish a
translator that can convert the concrete syntax of the theorem prover
into that of the proof checker. It may export explicit certificates
given by the \verb+ship+ tactic or derive certificates automatically, possibly
making use of proof scripts and execution traces in the theorem
file. These translation facilities may be built into the theorem
prover itself, or an instrumented version of it, thus encapsulating
the first two stages.  The translator outputs translation files usable
directly by a general-purpose, universal proof checker.

\parag{The proof checker.} The proof checker, as described in Section
\ref{sec:checker}, implements a focused version of the $\mu LJ$ logic
in the Bedwyr model checking system, augmented to further implement
the FPC framework. The translated theorems and their certificates plug
into this kernel and constitute a full instantiation of the system,
which Bedwyr can execute to verify that the certificates can be
elaborated into complete proofs of their associated theorems.

Translators are the only addition needed to enable a new theorem
prover to use the FPC framework.  Such translators are free to
implement sophisticated mechanisms to produce efficient certificates
from proof scripts but, in fact, only a more shallow syntactic
translation is strictly required: in this latter case, sophistication
must be built into the FPC definitions.  For the
present study with the Abella interactive theorem prover, the
translator has been integrated in an instrumented version of Abella, a
system whose proximity with Bedwyr syntax is naturally well-suited to
the approach.

The work-flow structure just described makes no explicit reference to
the FPCs that could be shipped, constructed, and checked.  These can
conform to the definition of proof outlines used throughout this paper
or be tailored to each specific problem.  The translator module can
choose to use proof outlines as the default, as is the case in our
examples. It could also let a user of \verb+ship+ specify an FPC
definition to be included in the resulting translation, or generate
tentative certificates with rules involving other FPC definitions.

\section{Conclusion}

We have described a methodology for elaborating proof outlines
using a framework for checking proof certificates.
We have illustrated this approach with ``simple inductive proofs''
that are applicable in a wide range of situations involving
inductively defined datatypes.
%
% Since our method is a direct implementation of simple proof theory
% concepts and since those proof theory concepts are also known to work
% for co-induction and for the $\lambda$-tree approach to syntax
% (including the $\nabla$-quantifier \cite{baelde14jfr}) we
% will be able to generalize this work to that richer setting.
%
We plan to scale and study significantly more complex examples in the
near future.
Finally, it would be interesting to see how our use of high-level
descriptions of proofs and proof reconstruction might be related to
the work of Bundy and his colleagues on proof plans and rippling
\cite{bundy87cade,bundy93ai}.

\smallskip
\noindent{\em Acknowledgments:} We thank anonymous reviewers
and K. Chaudhuri for comments on a draft of this paper.  This work was
funded by the ERC Advanced Grant Proof\kern 0.6pt Cert.

\bibliographystyle{eptcs}
\bibliography{../references/master}
\end{document}

%% file: listing-macros.tex
\colorlet{lprolog}{blue!70!black}